\documentclass[11pt,a4paper]{article}

\usepackage{CJKutf8}

\usepackage{mathtools}
\usepackage{authblk} 
\usepackage{algpseudocode,algorithmicx,algorithm}

\usepackage{mathrsfs}
\usepackage{latexsym,bm}
\usepackage{amsmath,amsfonts,amsmath,amssymb,amsthm}
\usepackage{extarrows}

\usepackage{graphicx,subfigure,epstopdf,float}
\usepackage{enumerate,cases,multirow}
\usepackage{caption}

\usepackage{longtable,colortbl,arydshln,threeparttable}
\definecolor{mygray}{gray}{.9}

\usepackage{indentfirst}
\usepackage[top=25mm,bottom=20mm,left=25mm,right=20mm]{geometry}
\usepackage{cite}

\usepackage{listings}

\usepackage{makeidx}        
\usepackage{booktabs}
\usepackage[unicode=true,bookmarksnumbered,colorlinks,citecolor=red,linkcolor=red,hyperindex,linktocpage=true]{hyperref}

\newcommand{\bb}{\boldsymbol}

\def \e {\mathrm{e}}
\def \i {\mathrm{i}}

\newcounter{parentalgorithm}

\makeatother

\newtheorem{theorem}{Theorem}[section]

\theoremstyle{remark}

\numberwithin{equation}{section}
\begin{document}
\begin{CJK*}{UTF8}{gbsn}

\title{Time complexity analysis of quantum difference methods for the multiscale transport equations}
\author{Xiaoyang He\thanks{hexiaoyang@sjtu.edu.cn}}
\author{Shi Jin\footnote{Corresponding author.} \thanks{shijin-m@sjtu.edu.cn}}
\author{Yue Yu\thanks{terenceyuyue@sjtu.edu.cn}}
\affil{School of Mathematical Sciences, Institute of Natural Sciences, MOE-LSC, Shanghai Jiao Tong University, Shanghai, 200240, P. R. China.}
\date{}
\maketitle

\begin{abstract}
  We investigate time complexities of finite difference methods for solving the multiscale transport equation with quantum algorithms. We find  that the time complexities of both the classical treatment and quantum treatment for a standard explicit scheme scale as $\mathcal{O}(1/\varepsilon)$, where $\varepsilon$ is the small scaling parameter,  while the complexities for the even-odd parity based Asymptotic-Preserving (AP) scheme do not depend on $\varepsilon$. This indicates that it is still of great importance to use AP (and probably  other efficient multiscale) schemes for multiscale problems in quantum computing when solving  multiscale transport or kinetic equations.
\end{abstract}


\clearpage
\tableofcontents

\section{Introduction}


Transport equations arise in many important applications, from medical imaging, astrophysics, nuclear reactor, to wave propagation in random media and semiconductor device modeling \cite{CaseZweifel,Chander,Ryzhik,MRS}.  These equations model probability distribution of particles in a background medium, thus are defined in phase space, suffering from curse-of-dimensionality.
In addition, the problem may encounter multiple temporal and spatial scales, and the numerical resolution of the small scales will further increase  the computational cost tremendously.
Despite of rapid development of multiscale methods, high dimensionality and multiple scales could still pose a major challenge for numerical simulations for transport, and more generally, kinetic equations by classical computers.

On the other hand, quantum computers, in various instances, have been shown to exhibit  potential  polynomial and even exponential advantage over the classical computers, if one designs adequate quantum algorithms. One of such possibilities is  linear algebra problems \cite{HHL-2009,Childs-2017,Gilyen2019QSVD}.  After numerical discretizations,  ordinary and partial differential equations can also be formulated as linear algebra problems thus can also use quantum linear algebra solvers to gain quantum advantages  in dimension, precision, and the number of simulations
\cite{Berry-2014,Joseph2020KvN,Dobin2021plasma,Lloyd2020nonlinear,Childs-2021,Liu2021nonlinear,JLY2022multiscale,JinLiu2022nonlinear}. Most of these works aim at producing quantum state, after which a measurement is needed to extract classical data. In \cite{Liu2021nonlinear} though, physical observables are obtained with possible quantum advantage.

In particular, in \cite{JLY2022multiscale}, for a linear hyperbolic relaxation system with possibly stiff relaxation, it shows that a good multiscale scheme~--~in this case the Asymptotic-preserving (AP) scheme  \cite{JPT1998,Jin2022Acta}~--~which is popular in kinetic community~--~has shown its advantage, for quantum algorithms, over standard non-AP schemes. Specifically, the numerical complexity that depends on the reciprocal of the small physically scaling scales is great relaxed: the complexity of AP quantum algorithms is {\it independent } of the small scaling parameter.

In this article we study the
multiscale linear transport equation
\begin{equation}\label{equ:1}
\epsilon \partial_{t} f+v \partial_x f
=\frac{1}{\epsilon}\left(\frac{1}{2} \int_{-1}^{1} f d v^{\prime}-f\right), \quad x_{L}<x<x_R, \quad -1\leq v\leq 1,
\end{equation}
where $f=f(t,x,v)$ is the probability density distribution for particles at space point $x \in \mathbb{R}$, time $t$, and  $v\in (-1,1)$ is the cosine of the angle between the particle velocity and the $x$-axis. Comparing with the work in \cite{JLY2022multiscale}, here the equation is in the phase space, and one needs to also discretize  the velocity (or angle) variable, and to deal with the nonlocal collision operator, hence further complicating the development of numerical approximations and the study of their time complexity for quantum algorithms. Our goal is to compare the time complexity of quantum algorithms based on an AP scheme \cite{JPT1998} and a standard (explicit, thus not AP) scheme and show that the former has a complexity {\it independent of} the small physical scaling parameter $\epsilon$ while the latter depends on it. Hence it demonstrates that  multiscale methods still
make a big difference in terms of time compexity  even for quantum algorithms.

Since our aim is to compare the difference in dependence of $\epsilon$, in this article we  will only study the spatially one dimensional equation. Quantum advantages in spatial dimensions for numerical methods of partial differential equations have been well studied in other literature, see for examples \cite{Childs-2021,Lin-Montanaro-Shao-2020, JLY2022multiscale, JinLiuYu-2}


Compared with the earlier work \cite{JLY2022multiscale} where a multiscale hyperbolic relaxation system was studied, here in the time complexity analysis for transport equation defined in the phase space with a nonlocal collisional term,   the analytic difficulty  is to give a lower bound of the minimum singular value of the coefficient matrix. When neglecting the nonlocal term, one easily observes that the problem is reduced to the prototype problem for fixed velocity variable. Its simplicity allows one to estimate the singular values of the coefficient matrix directly, in which the proof ultimately boils down to the upper bound of the 2-norm of the inverse matrix of $\bb{K}$ in the form of
\[\bb{K} =
\begin{bmatrix}
\bb{I}          &                  &            &               \\
-\bb{B}_1       &  \bb{I}          &            &              \\
                &  \ddots          &  \ddots    &                \\
                &                  &  -\bb{B}_1 &  \bb{I}    \\
\end{bmatrix},\]
which satisfies $\|\bb{B}_1\|\le 1$ under an approriate CFL condition and leads to the expected estimate
\[\|\bb{K}^{-1}\| \le 1 + \|\bb{B}_1\| + \|\bb{B}_1\|^2 + \cdots + \|\bb{B}_1\|^{N_t-1} \le N_t.\]
In contrast, the inclusion of the integral  term due to the nonlocal collision operator makes the discretization {\it fully coupled} in the angular direction, so the direct manipulation of the coefficient matrix will be rather involved as opposed to the analysis for the prototype problem. For this reason, we instead characterize the singularity by using the Fourier analysis approach on the spatial variable, which enables us to derive the CFL condition quite naturally, and makes the system more convenient to perform the perturbation technique. Our analysis also relies on the special properties of a rank-one matrix  composed of the weights of the numerical integration as introduced in the proof of Theorem \ref{the:3.1}.

\section{The even-odd parity based AP scheme for the multiscale transport equation}

In this section we will review an AP scheme (the diffusive relaxation scheme), proposed in \cite{JPT1998}, for \eqref{equ:1}.

\subsection{The diffusive relaxation system}

The transport equation can be reformulated to the diffusive relaxation system. To this end, let us split it into two equations, each for $v>0$:

\begin{equation}
    \begin{cases}
        \label{equ:2}
        \epsilon \partial_{t} f(v)+v \partial_x f(v) =\frac{1}{\epsilon}\left(\frac{1}{2} \int_{-1}^{1} f d v-f(v)\right), \\
        \epsilon \partial_{t} f(-v)-v \partial_x f(-v) =\frac{1}{\epsilon}\left(\frac{1}{2} \int_{-1}^{1} f d v-f(-v)\right) .
    \end{cases}
\end{equation}
Introducing the even- and odd-parities
\begin{equation}
    \begin{cases}
        \label{equ:3}
        r(t, v, x)=\frac{1}{2}[f(t, v, x)+f(t, -v,x)], \\
        j(t, v, x)=\frac{1}{2 \epsilon}[f(t, v, x)-f(t, -v,x)] ,
    \end{cases}
\end{equation}
one has the following system
\begin{equation}
    \begin{cases}
        \label{equ:5}
        \partial_{t} r+v \partial_x j =\frac{1}{\epsilon^2}(\rho-r), \\
        \partial_{t} j+\frac{v}{\epsilon^2} \partial_x r =-\frac{1}{\epsilon^2} j,
    \end{cases}
\end{equation}
where
\begin{equation*}
        \rho(t, x)=\int_{0}^{1} r d v.
\end{equation*}
The idea of \cite{Jin-Pareschi-Toscani-2000} is to rewrite \eqref{equ:5} as the following diffusive relaxation system
\begin{equation}
    \begin{cases}
        \label{equ:6}
        \partial_{t} r+v \partial_x j =-\frac{1}{\epsilon^2}(r-\rho) ,\\
        \partial_{t} j+\phi v \partial_x r =-\frac{1}{\epsilon^2}\left(j+\left(1-\epsilon^2 \phi\right) v \partial_x r\right),
    \end{cases}
\end{equation}
where $\phi=\phi(\epsilon)$ satisfies $0 \leq \phi \leq 1 / \epsilon^{2}$. The requirement of $\phi$ guarantees that $\phi(\epsilon)$ and $ 1-\epsilon^2 \phi (\epsilon)$ are positive, making the problem uniformly stable
when $\epsilon$ is small. A simple choice is $\phi(\epsilon)=\min \{1, 1/\epsilon\}$.
In what follows, we take $\phi = 1$ since we are mainly concerned with the case of $\epsilon \ll 1$.\\

\subsection{The diffusive relaxation scheme}

Ref.~\cite{Jin-Pareschi-Toscani-2000} presented a natural splitting of ~\eqref{equ:6}~ which consists of combining the relaxation step
\begin{equation}
    \begin{cases}
        \label{equ:7}
        \partial_{t} r =-\frac{1}{\epsilon^2}(r-\rho), \\
        \partial_{t} j =-\frac{1}{\epsilon^2}\left(j+\left(1-\epsilon^2 \phi\right) v \partial_x r\right),
    \end{cases}
\end{equation}
with the transport step
\begin{equation}
    \begin{cases}
        \label{equ:8}
        \partial_{t} r+v \partial_x j=0, \\
        \partial_{t} j+\phi v \partial_x r=0.
    \end{cases}
\end{equation}
Now we use some discretization methods to deal with the two steps.

\subsubsection{The relaxation step}
To have a good stability property, considering the implicit discretization for the relaxation term \eqref{equ:7}, one can obtain
\begin{equation} \label{equ:9}
\begin{cases}
\frac{r^*-r^n}{\tau}=-\frac{1}{\epsilon^2}(r^*-\rho^*), \\
\frac{j^*-j^n}{\tau}=-\frac{1}{\epsilon^2}\Big(j^*+(1-\epsilon^2 \phi) v \partial_x r^*\Big) ,
\end{cases}
\end{equation}
We remark that the above system can be implemented explicitly on a classical computer since $\rho$ is preserved , i.e., $\rho^*=\rho^n$, and hence
\begin{equation}
\begin{cases}
\frac{r^*-r^n}{\tau}=-\frac{1}{\epsilon^2}(r^*-\rho^n), \\
\frac{j^*-j^n}{\tau}=-\frac{1}{\epsilon^2}\Big(j^*+ (1-\epsilon^2 \phi ) v \partial_x r^*\Big),
\end{cases}
\end{equation}
where the integral given by $\rho$ will be approximated by the Gaussian quadrature rule
\begin{equation*}
\rho(t,x) = \int_0^1 r(t,v,x) {\rm d}v \approx \sum_{k=1}^N w_kr(t,v_k,x),
\end{equation*}
with $(v_k, w_k)$ being the Gaussian quadrature points and weights on $[0,1]$.

The spatial mesh of $x$ is defined as $x_0<x_1<\cdots<x_{N_x}<x_{N_x+1}$ (where $x_0$ and $x_{N_x+1}$ are boundary points), and the discrete time are $t_0<t_1<\cdots<t_{N_t}$.
Let $u_{km}$ be the approximation to $u(v_k,x_m)$. Then the discrete scheme of \eqref{equ:9} is
\begin{equation}\label{relaxation}
\begin{cases}
r_{km}^*=\frac{1}{1+\tau/\epsilon^2}(r_{km}^n+\frac{\tau}{\epsilon^2}\rho_m^n),\\
j_{km}^*=\frac{1}{1+\tau/\epsilon^2}\Big(j_{km}^n-\frac{\tau}{\epsilon^2}(1-\epsilon^2)v_k\frac{r_{k,m+1}^*-r_{k,m-1}^*}{2h} \Big),
\end{cases}
\end{equation}
where~$k=1,\cdots,N$ and $m = 1,\cdots,N_x$.
For fixed $v_k$, we define $\bb{u}_k = [u_{k1}, u_{k2}, \cdots,u_{k,N_x}]^T$ and write \eqref{relaxation} in vector form as
\begin{equation*}\label{rkjk1}
\begin{cases}
\bb{r}_k^*=\frac{1}{1+\tau/\epsilon^2}\Big( \bb{r}_k^n +  \frac{\tau}{\epsilon^2} (w_1 \bb{r}_1^n + \cdots + w_N \bb{r}_N^n) \Big) \\
\bb{j}_k^* = \frac{1}{1+\tau/\epsilon^2} \Big( \bb{j}_k^n - \frac{\tau}{2h \epsilon^2}(1-\epsilon^2)v_k\bb{M}_h\bb{r}_k^* - \frac{\tau}{2h \epsilon^2}(1-\epsilon^2)v_k \widetilde{\bb{b}}_k^* \Big)
\end{cases}, \quad k = 1,2,\cdots, N,
\end{equation*}
where,
\[\bb{M}_h =
\begin{bmatrix}
0    &  1        &                &    &       \\
 -1  &  0        &  \ddots        &    &        \\
     & \ddots    &   \ddots       & \ddots  &   \\
     &           &    \ddots         & 0  &    1    \\
     &           &                &  -1  &    0
\end{bmatrix}_{N_x \times N_x}, \qquad
\widetilde{\bb{b}}_k(t) =
        \begin{bmatrix}
        -r_{k0}(t) \\
        0  \\
        \vdots\\
        0 \\
        r_{k,N_x+1}(t)
        \end{bmatrix}.
\]
Let $\bb{r} = [\bb{r}_1; \bb{r}_2; \cdots; \bb{r}_N]$,  where ``;" indicates the straightening of $\{\bb{r}^i\}_{i\ge 1}$ into a column vector. Then,
\begin{equation*}
\begin{cases}
\bb{r}^* = \frac{1}{1+\tau/\epsilon^2} \Big(  \bb{I} +  \frac{\tau}{\epsilon^2} \bb{G} \Big)\bb{r}^n, \\
\bb{j}^* = \frac{1}{1+\tau/\epsilon^2}\Big(  \bb{j}^n -\frac{\tau}{2h\epsilon^2}(1-\epsilon^2) \bb{M}_v\bb{r}^*
        -\frac{\tau}{2h\epsilon^2}(1-\epsilon^2)\widetilde{\bb{b}}_v^*\Big),
\end{cases}
\end{equation*}
where $\widetilde{\bb{b}}_v^*= [v_1\widetilde{\bb{b}}_1;\cdots; v_N\widetilde{\bb{b}}_N ]$, and
\[\bb{G} = \begin{bmatrix}
        w_1 \bb{I}  & w_2 \bb{I} & \cdots & w_N \bb{I} \\
        w_1 \bb{I}  & w_2 \bb{I} & \cdots & w_N \bb{I} \\
        \vdots      & \vdots     & \ddots & \vdots     \\
        w_1 \bb{I}  & w_2 \bb{I} & \cdots & w_N \bb{I} \\
        \end{bmatrix}_{NN_x\times NN_x }, \quad
\bb{M}_v = \begin{bmatrix}
        v_1 \bb{M}_h  &               &         &  \\
                      & v_2 \bb{M}_h  &         &   \\
                      &               & \ddots  &    \\
                      &               &         & v_N \bb{M}_h \\
        \end{bmatrix}.
\]

\subsubsection{The transport step}

For the transport step~\eqref{equ:8}, by introducing the Riemann invariants $U=r+\phi^{-1 / 2} j$ and $V=r-\phi^{-1 / 2} j$, one can obtain
\begin{equation*}
\begin{cases}
\partial_{t} U+\phi^{1 / 2} v \partial_x U=0, \\
\partial_{t} V-\phi^{1 / 2} v \partial_x V=0.
\end{cases}
\end{equation*}
Applying the  upwind scheme to the spatial derivative gives
\begin{equation}\label{diffusion}
\begin{cases}
r_{km}^{n+1}=(1-\lambda v_k)r_{km}^*+\frac{\lambda v_k}{2}(r_{k,m+1}^*+r_{k,m-1}^*)-\frac{\lambda v_k}{2}(j_{k,m+1}^*-j_{k,m-1}^*),\\
j_{km}^{n+1}=(1-\lambda v_k)j_{km}^*+\frac{\lambda v_k}{2}(j_{k,m+1}^*+j_{k,m-1}^*)-\frac{\lambda v_k}{2}(r_{k,m+1}^*-r_{k,m-1}^*),
\end{cases}
\end{equation}
where~$\lambda=\tau/h$ and $\phi =1$.
Similarly, for the fixed~$k$, one has
\begin{equation*}\label{rkjk2}
\begin{cases}
\bb{r}_k^{n+1}=(\bb{I} + \frac{\lambda v_k}{2} \bb{L}_h)\bb{r}_k^*-\frac{\lambda v_k}{2}\bb{M}_h \bb{j}_k^*
   + \frac{\lambda v_k}{2} (\bb{b}_k^*-\widetilde{\bb{c}}_k^*)\\
\bb{j}_k^{n+1}=(\bb{I} + \frac{\lambda v_k}{2} \bb{L}_h)\bb{j}_k^*-\frac{\lambda v_k}{2}\bb{M}_h \bb{r}_k^*
   + \frac{\lambda v_k}{2} (\bb{c}_k^*-\widetilde{\bb{b}}_k^*)
\end{cases}, \quad k = 1,2,\cdots, N,
\end{equation*}
where,
\[\bb{L}_h =\begin{bmatrix}
        -2  &  1       &                &      &         \\
         1  & -2       & \ddots         &      &          \\
            &  \ddots  & \ddots         & \ddots &         \\
            &          &   \ddots       &      &  1  \\
            &          &                &1     & -2 \\
\end{bmatrix}, \quad
    \begin{aligned}
        \bb{b}_k =
        \begin{bmatrix}
        r_{k0} \\
        0  \\
        \vdots\\
        0 \\
        r_{k,N_x+1} \\
        \end{bmatrix}, \quad
        \bb{c}_k =
        \begin{bmatrix}
        j_{k0} \\
        0  \\
        \vdots\\
        0 \\
        j_{k,N_x+1} \\
        \end{bmatrix}, \quad
        \widetilde{\bb{c}}_k =
        \begin{bmatrix}
        -j_{k0} \\
        0  \\
        \vdots\\
        0 \\
        j_{k,N_x+1} \\
        \end{bmatrix}.
    \end{aligned}
\]
We rewrite the scheme in matrix form:
\begin{equation*}
\begin{cases}
 \bb{r}^{n+1} =  ( \bb{I}+ \frac{\lambda}{2} \bb{L}_v ) \bb{r}^* - \frac{\lambda}{2}\bb{M}_v\bb{j}^*
                    + \frac{\lambda}{2} \bb{f}_v^*, \\
\bb{j}^{n+1} = ( \bb{I} + \frac{\lambda}{2} \bb{L}_v ) \bb{j}^* - \frac{\lambda}{2} \bb{M}_v\bb{r}^*
                    + \frac{\lambda}{2} \bb{g}_v^*,
 \end{cases}
\end{equation*}
where
\begin{equation*}
\bb{L}_v = \begin{bmatrix}
v_1 \bb{L}_h  &                &  \\
              &        \ddots  &    \\
              &                & v_N \bb{L}_h \\
 \end{bmatrix}, \qquad \bb{f}_v^* = \begin{bmatrix}
 v_1 (\bb{b}_1^* -\widetilde{\bb{c}}_1^*)\\
 \vdots \\
 v_N(\bb{b}_N^*-\widetilde{\bb{c}}_N^*)
 \end{bmatrix}, \quad \bb{g}_v^* = \begin{bmatrix}
 v_1(\bb{c}_1^*-\widetilde{\bb{b}}_1^*)\\
 \vdots \\
 v_N(\bb{c}_N^*-\widetilde{\bb{b}}_N^*)
 \end{bmatrix}.
\end{equation*}

\section{Time complexity analysis of the AP scheme}

\subsection{The quantum difference method}

Let $\bb{A}=\frac{\lambda}{2}\bb{M}_v$, $\bb{B}=\bb{I}+ \frac{\lambda}{2} \bb{L}_v$ and $\gamma = \tau/\epsilon^2$. Substituting $\bb{r}^*$ and $\bb{j}^*$ into $\bb{r}^{n+1}$, we obtain
\begin{align*}
\bb{r}^{n+1}
& = \bb{B} \bb{r}^*-\bb{A} \bb{j}^*+\frac{\lambda}{2}\bb{f}_v^*\\
& = \bb{B}\bb{r}^*-\frac{1}{1+\gamma}\bb{A}\Big(\bb{j}^n-\frac{(1-\epsilon^2)}{\epsilon^2}\bb{A} \bb{r}^*-\frac{\lambda(1-\epsilon^2)}{2\epsilon^2}\widetilde{\bb{b}}_v^*\Big)+\frac{\lambda}{2}\bb{f}_v^*\\
& = \Big(\bb{B} + \frac{1}{1+\gamma}\frac{(1-\epsilon^2)}{\epsilon^2}\bb{A}^2\Big)\bb{r}^*
    - \frac{1}{1+\gamma} \bb{A}\bb{j}^n + \frac{1}{1+\gamma}\frac{\lambda(1-\epsilon^2)}{2\epsilon^2} \bb{A}\widetilde{\bb{b}}_v^* + \frac{\lambda}{2}\bb{f}_v^*\\
& = \frac{1}{1+\gamma} \Big(\bb{B} + \frac{1}{1+\gamma}\frac{(1-\epsilon^2)}{\epsilon^2}\bb{A}^2\Big)(\bb{I}+\gamma \bb{G}) \bb{r}^n
    - \frac{1}{1+\gamma} \bb{A}\bb{j}^n  \\
& \quad + \frac{1}{1+\gamma}\frac{\lambda(1-\epsilon^2)}{2\epsilon^2} \bb{A}\widetilde{\bb{b}}_v^* + \frac{\lambda}{2}\bb{f}_v^*\\
& = \frac{1}{1+\gamma} \Big(\bb{B} + \frac{1-\epsilon^2}{\tau+\epsilon^2}\bb{A}^2\Big)(\bb{I}+\gamma \bb{G}) \bb{r}^n
    - \frac{1}{1+\gamma} \bb{A}\bb{j}^n + \frac{\lambda(1-\epsilon^2)}{2(\tau + \epsilon^2)} \bb{A}\widetilde{\bb{b}}_v^* + \frac{\lambda}{2}\bb{f}_v^*\\
& =: \bb{B}_1\bb{r}^n - \bb{A}_1 \bb{j}^n + \widetilde{\bb{f}}^{n+1},
\end{align*}
where,
\[\bb{B}_1 = \frac{1}{1+\gamma} \Big(\bb{B} + \frac{1-\epsilon^2}{\tau+\epsilon^2}\bb{A}^2\Big)(\bb{I}+\gamma \bb{G}), \qquad
\bb{A}_1 = \frac{1}{1+\gamma} \bb{A}.\]
A similar calculation gives
\begin{align*}
\bb{j}^{n+1}
& = \bb{B}\bb{j}^* - \bb{A} \bb{r}^* + \frac{\lambda}{2}\bb{g}_v^*\\
& = \frac{1}{1+\gamma} \bb{B}\Big(\bb{j}^n-\frac{(1-\epsilon^2)}{\epsilon^2}\bb{A} \bb{r}^*-\frac{\lambda(1-\epsilon^2)}{2\epsilon^2}\widetilde{\bb{b}}_v^*\Big)-\bb{A}\bb{r}^*+\frac{\lambda}{2}\bb{g}_v^*\\
& = \frac{1}{1+\gamma}\bb{B}\bb{j}^n -\Big(  \frac{1}{1+\gamma}\frac{1-\epsilon^2}{\epsilon^2} \bb{B}\bb{A} + \bb{A}\Big)\bb{r}^*
  -\frac{1}{1+\gamma} \frac{\lambda(1-\epsilon^2)}{2\epsilon^2}  \bb{B} \widetilde{\bb{b}}_v^*
        +\frac{\lambda}{2}\bb{g}_v^* \\
& = \frac{1}{1+\gamma}\bb{B}\bb{j}^n
  -\frac{1}{1+\gamma}\Big(\frac{1-\epsilon^2}{\tau+\epsilon^2} \bb{B}\bb{A} + \bb{A}\Big)(\bb{I}+\gamma\bb{G})\bb{r}^n
  -\frac{\lambda(1-\epsilon^2)}{2(\tau+\epsilon^2)}  \bb{B} \widetilde{\bb{b}}_v^*
        +\frac{\lambda}{2}\bb{g}_v^* \\
& =: \bb{A}_2 \bb{j}^n - \bb{B}_2 \bb{r}^n + \widetilde{\bb{g}}^{n+1},
\end{align*}
where,
\[\bb{B}_2 = \frac{1}{1+\gamma}\Big(\bb{A} + \frac{1-\epsilon^2}{\tau+\epsilon^2} \bb{B}\bb{A}\Big)(\bb{I}+\gamma\bb{G}), \qquad
\bb{A}_2 = \frac{1}{1+\gamma}\bb{B}.\]

Introducing the following notations
\[\bb{S}_1=[\bb{r}^1;\bb{r}^2;\cdots;\bb{r}^{N_t}],  \quad \bb{S_2}=[\bb{j}^1;\bb{j}^2;\cdots;\bb{j}^{N_t}],  \quad \bb{S}=[\bb{S_1};\bb{S_2}],\]
one obtains the linear system
\begin{equation}\label{equ:10}
\bb{L S}=\bb{F},
\end{equation}
where $\bb{L}=(\bb{L}_{i j})_{2 \times 2}$ and $\bb{F}=[\bb{F}_1; \bb{F}_2]$, with
\begin{equation*}
    \bb{L}_{11} =
    \begin{bmatrix}
    \bb{I}    &           &                &          \\
    -\bb{B}_1 &  \bb{I}   &                &          \\
              & \ddots    &   \ddots       &          \\
              &           &     -\bb{B}_1  & \bb{I}   \\
    \end{bmatrix},
        \bb{L}_{12} =
    \begin{bmatrix}
    \bb{O}    &           &                &          \\
     \bb{A}_1 &  \bb{O}   &                &          \\
              & \ddots    &   \ddots       &          \\
              &           &      \bb{A}_1  & \bb{O}   \\
    \end{bmatrix},
        \bb{F}_{1} =
    \begin{bmatrix}
    \widetilde{\bb{f}}^1+\bb{B}_1\bb{r}^0-\bb{A}_1\bb{j}^0\\
    \widetilde{\bb{f}}^2\\
    \vdots\\
    \widetilde{\bb{f}}^{N_t}\\
    \end{bmatrix},
\end{equation*}

\begin{equation*}
    \bb{L}_{21} =
    \begin{bmatrix}
    \bb{O}    &           &                &          \\
    \bb{B}_2 &  \bb{O}   &                &          \\
              & \ddots    &   \ddots       &          \\
              &           &     \bb{B}_2  & \bb{O}   \\
    \end{bmatrix},
        \bb{L}_{22} =
    \begin{bmatrix}
    \bb{I}    &           &                &          \\
    -\bb{A}_2 &  \bb{I}   &                &          \\
              & \ddots    &   \ddots       &          \\
              &           &     -\bb{A}_2  & \bb{I}   \\
    \end{bmatrix},
        \bb{F}_{2} =
    \begin{bmatrix}
    \widetilde{\bb{g}}^1-\bb{B}_2\bb{r}^0+\bb{A}_2\bb{j}^0\\
    \widetilde{\bb{g}}^2\\
    \vdots\\
    \widetilde{\bb{g}}^{N_t}\\
    \end{bmatrix}.
\end{equation*}

For fixed step sizes $\tau$ and $h$, when~$\epsilon \to 0$, one has
\[\frac{1}{1+\gamma} \to 0,\qquad \frac{\gamma}{1+\gamma} \to 1,\]
and hence
\[\bb{B}_2 = \Big(\bb{A} + \frac{1-\epsilon^2}{\tau+\epsilon^2} \bb{B}\bb{A}\Big)\frac{1}{1+\gamma}(\bb{I}+\gamma\bb{G}) \to
\Big(\bb{A} + \frac{1}{\tau} \bb{B}\bb{A}\Big) \bb{G} .\]
Considering the amplification factor $\tau^{-1}$,  as in
\cite{JLY2022multiscale} we reformulate the linear system as
\begin{equation}\label{reformulation}
\begin{bmatrix}
\bb{L}_{11} & \tau^{-1} \bb{L}_{12} \\
\tau \bb{L}_{21} & \bb{L}_{22}
 \end{bmatrix}
 \begin{bmatrix}
\tau^{-1} \bb{S}_1 \\
\bb{S}_2
 \end{bmatrix}
 = \begin{bmatrix}
\tau^{-1} \bb{F}_1 \\
\bb{F}_2
 \end{bmatrix},
\end{equation}
where $\tilde{\bb{S}}_1 = \tau^{-1} \bb{S}_1$ and $\tilde{\bb{S}}_2 = \bb{S}_2$. This means we consider a linear system with new variables $\tilde{r} = \tau^{-1} r$ and $\tilde{j} = j$.

\subsection{The time complexity of the AP scheme}

In this article we apply the optimal quantum linear systems algorithm (QLSA) proposed in
\cite{Costa2021QLSA} to solve the resulting linear system. The query complexity with respect to the sparse access to matrices can be written as
\begin{equation}\label{cpCAS}
Q = \mathcal{O} ( s \kappa \log(1/\delta)   ),
\end{equation}
where $s$ is the sparsity of the coefficient matrix, $\kappa$ is the condition number, and $\delta$ is the error bound.

For input models of querying the matrix and vector, we refer the reader to \cite{BerryChilds2017ODE,JinLiu2022nonlinear,Low-2019,JLY2022multiscale,
gleinig2021efficient,zhang2022quantum}.

\begin{theorem}\label{the:3.1}
Suppose that the time step $\tau$ and the space step $h$ satisfy $\tau/h^2 \le 1/(1+h)$.
\begin{enumerate}[(1)]
\item For sufficiently small~$\epsilon$,
the singular value of the coefficient matrix in~\eqref{reformulation}~satisfies
\[\sigma_{\min} \gtrsim  1/(N^{1/2}N_t) - \alpha(\epsilon), \qquad \sigma_{\max} \lesssim N^{1/2}+ \alpha(\epsilon),\]
where
\[
\alpha(\epsilon)=\frac{\epsilon^2}{\tau+\epsilon^2}(N^{1/2}\tau+N^{1/2}+\tau+\frac{1}{\tau})
+\frac{\epsilon^2(1-\epsilon^2)}{(\epsilon^2+\tau)^2}(1+\tau)
+\frac{\epsilon^2(\epsilon^2+2\tau+\tau^2)}{\tau(\epsilon^2+\tau)^2}N^{1/2}(1+\frac{1}{\tau}),
\]
which tends to zero as $\epsilon \to 0$.
  \item The sparsity and the condition number of the coefficient matrix satisfy $s = \mathcal{O}(N)$ and $\kappa = \mathcal{O}(NN_t)$.
 \item  The time complexities of the classical treatment and the quantum treatment for solving \eqref{reformulation} are
\[C = \mathcal{O}(N^2 N_t N_x),\qquad Q = \mathcal{O}(N^2 N_t \log(N_x)).\]
If~$N_t = \mathcal{O}(N_x^2)$, then
\[C = \mathcal{O}(N^2 N_x^3),\qquad Q = \mathcal{O}(N^2 N_x^2 \log(N_x)).\]
\end{enumerate}
\end{theorem}
\begin{proof}
Since the problem is linear, one can apply the discrete Fourier transform to
characterize the singular values of the coefficient matrix. In the following, we only consider the discrete Fourier transform for the spatial variables.

1) Introduce the following expressions
\[
r_{km}^n=\hat{r}_k^n \e^{\i m\xi h}, \quad j_{km}^n=\hat{j}_k^n \e^{\i m\xi h}, \quad
 r_{km}^*=\hat{r}_k^* \e^{\i m \xi h}, \quad  j_{km}^*=\hat{j}_k^* \e^{\i m\xi h},
\]
where $\xi$ represents the frequency variable and $\i=\sqrt{-1}$. Plugging them in \eqref{relaxation} and~\eqref{diffusion}, one obtains
\begin{align*}
&\hat{r}_k^*=\frac{1}{1+\gamma}(\hat{r}_k^n+\gamma\sum_{k'=1}^{N}w_{k'}\hat{r}_{k'}^n),\\
&\hat{j}_k^*=\frac{1}{1+\gamma}\Big(\hat{j}_k^n- \frac{1-\epsilon^2}{\epsilon^2}  \i \lambda \sin(\xi h) v_k\hat{r}_k^* \Big)
\end{align*}
and
\begin{align*}
& \hat{r}_k^{n+1}=(1-\lambda v_k)\hat{r}_k^*+  \lambda \cos(\xi h) v_k \hat{r}_k^*-\i \lambda \sin(\xi h) v_k \hat{j}_k^*,\\
& \hat{j}_k^{n+1}=(1-\lambda v_k)\hat{j}_k^*+\lambda \cos(\xi h) v_k\hat{j}_k^*-\i \lambda \sin(\xi h) v_k \hat{r}_k^*.
\end{align*}
Eliminating $r^*$ and $j^*$ yields
\begin{equation}\label{equ:F1}
    \begin{cases}
    \hat{r}_k^{n+1}+c_{1,\epsilon}\hat{r}_k^{n}+c_{2,\epsilon}\hat{j}_k^n
    +\gamma c_{1,\epsilon}\sum_{k'=1}^{N}w_{k'}\hat{r}_{k'}^n=0,\\
    \hat{j}_k^{n+1}+d_{1,\epsilon}\hat{j}_k^{n}+d_{2,\epsilon}\hat{r}_k^n
    +\gamma d_{2,\epsilon}\sum_{k'=1}^{N}w_{k'}\hat{r}_{k'}^n=0,
    \end{cases}
\end{equation}
where~$n=0,1,\cdots,N_t-1$ and
\begin{align*}
    & c_{1,\epsilon}=-\frac{1}{1+\gamma}[(1-\lambda v_k)+\lambda v_k \cos(\xi h)]
    -\frac{1}{(1+\gamma)^2}\frac{1-\epsilon^2}{\epsilon^2}(\i\lambda v_k \sin(\xi h))^2,\\
    & c_{2,\epsilon}=\frac{1}{1+\gamma}\i\lambda v_k \sin(\xi h),\\
    & d_{1,\epsilon}=-\frac{1}{1+\gamma}[(1-\lambda v_k)+\lambda v_k \cos(\xi h)]\\
    & d_{2,\epsilon}=\frac{1}{(1+\gamma)^2}\frac{1-\epsilon^2}{\epsilon^2}[(1-\lambda v_k)+\lambda v_k \cos(\xi h)]\i\lambda v_k\sin(\xi h)
    +\frac{1}{1+\gamma}\i\lambda v_k \sin(\xi h).
\end{align*}
For the new variables, the linear system \eqref{equ:F1} should be changed to
\begin{equation*}
    \begin{cases}
    \tau \tilde{r}_k^{n+1}+c_{1,\epsilon}\tau\tilde{r}_k^{n}+c_{2,\epsilon}\tilde{j}_k^n
    +\gamma c_{1,\epsilon}\tau\sum_{k'=1}^{N}w_{k'}\tilde{r}_{k'}^n=0,\\
    \tilde{j}_k^{n+1}+d_{1,\epsilon}\tilde{j}_k^{n}+d_{2,\epsilon}\tau \tilde{r}_k^n
    +\gamma d_{2,\epsilon}\tau \sum_{k'=1}^{N}w_{k'}\tilde{r}_{k'}^n=0.
    \end{cases}
\end{equation*}

Let~$\tilde{\bb{r}}_k=[\tilde{r}_k^{1}, \cdots, \tilde{r}_k^{N_{t}}]^T$,~$\tilde{\bb{j}}_k=[\tilde{j}_k^{1}, \cdots, \tilde{j}_k^{N_{t}}]^T$, and
\[P = \begin{bmatrix}
    0    &           &                &          \\
    1 &  0   &                &          \\
              & \ddots    &   \ddots       &          \\
              &           &     1  & 0   \\
    \end{bmatrix}_{N_t\times N_t}.\]
Then Eq.~\eqref{equ:F1} can be written as
\begin{align*}
\tau (I+c_{1,\epsilon}P)\tilde{\bb{r}}_k +c_{2,\epsilon} P \tilde{\bb{j}}_k
    + \tau \gamma c_{1,\epsilon}P (w_1 \tilde{\bb{r}}_1 + \cdots + w_N \tilde{\bb{r}}_N) = \tilde{\bb{f}}_k,\\
(I+d_{1,\epsilon}P)\tilde{\bb{j}}_k + \tau d_{2,\epsilon} P \tilde{\bb{r}}_k
    + \tau \gamma d_{2,\epsilon}P (w_1 \tilde{\bb{r}}_1 + \cdots + w_N \tilde{\bb{r}}_N) = \tilde{\bb{g}}_k,
\end{align*}
where~$k=1,2,\cdots,N$, and the right-hand vectors are
\[\tilde{\bb{f}}_k = -[\tau c_{1,\epsilon}\tilde{r}_k^0 + c_{2,\epsilon}\tilde{j}_k^0 + \tau \gamma c_{1,\epsilon}(w_1 \tilde{r}_1^0+\cdots+w_N\tilde{r}_N^0), 0, \cdots, 0]^T, \]
\[\tilde{\bb{g}}_k = -[d_{1,\epsilon}\tilde{j}_k^0 + \tau d_{2,\epsilon}\tilde{r}_k^0 + \tau \gamma d_{2,\epsilon}(w_1 \tilde{r}_1^0+\cdots+w_N\tilde{r}_N^0), 0, \cdots, 0]^T. \]
Let~$\tilde{\bb{R}}=[\tilde{\bb{r}}_1;\tilde{\bb{r}}_2;\cdots;\tilde{\bb{r}}_N]$, $\tilde{\bb{J}}=[\tilde{\bb{j}}_1;\tilde{\bb{j}}_2;\cdots;\tilde{\bb{j}}_N]$~and~$\tilde{\bb{S}}=[\tilde{\bb{R}};\tilde{\bb{J}}]$, one obtains the linear system
\begin{align}\label{linearsystemFourier}
\tilde{\bb{L}}_{\epsilon}\tilde{\bb{S}} = \tilde{\bb{F}},
\end{align}
where $\tilde{\bb{F}} = [\tilde{\bb{f}}_1 /\tau; \cdots; \tilde{\bb{f}}_N /\tau; \tilde{\bb{g}}_1; \cdots; \tilde{\bb{g}}_N]$,
\[\tilde{\bb{L}}_{\epsilon}=
    \begin{bmatrix}
    I_N \otimes (I+c_{1,\epsilon}P) + \gamma c_{1,\epsilon} W \otimes P    &    \tau^{-1} I_N \otimes (c_{2,\epsilon}P)      \\
    \tau ( I_N \otimes (d_{2,\epsilon}P)  + \gamma d_{2,\epsilon} W  \otimes P)    &    I_N \otimes (I+d_{1,\epsilon}P)    \\
    \end{bmatrix},\]
and
\[W = \begin{bmatrix}
w_1 &    w_2      &   \cdots      &  w_N   \\
w_1 &    w_2      &   \cdots      &  w_N   \\
\vdots &    \vdots      &   \ddots      &  \vdots   \\
w_1 &    w_2      &   \cdots      &  w_N   \\
\end{bmatrix}_{N\times N}.
\]

2) In the following, we utilize the perturbation technique to analyze the condition number of the coefficient matrix. We first briefly explain the idea of the perturbation technique: Let $\tilde{\bb{L}}_{\epsilon}$ be the coefficient matrix and $\tilde{\bb{L}}_{\epsilon}=\tilde{\bb{L}}_0+{\bb{E}}$, where $\tilde{\bb{L}}_0$ is the coefficient matrix with $\epsilon=0$. By the Weyl's inequality\cite{Stewart1990PerturbationTF},
\[
\sigma_{\max}(\tilde{\bb{L}}_{\epsilon})\le \sigma_{\max}(\tilde{\bb{L}}_0) + \|{\bb{E}}\|,\quad
\sigma_{\min}(\tilde{\bb{L}}_{\epsilon}) \ge
\sigma_{\min}(\tilde{\bb{L}}_0)- \|{\bb{E}}\|.
\]
Thus, it suffices to determine the condition number of $\tilde{\bb{L}}_0$ and the upper bound of $\|\boldsymbol{E}\|$.

Let $\epsilon=0$. One has
\[\frac{1}{1+\gamma_0}=0,\quad \frac{\gamma_0}{1+\gamma_0}=1,\]
where~$\gamma_0$~corresponds to~$\epsilon=0$. A simple calculation shows that
\[c_{1,0} = c_{2,0} = d_{1,0} = d_{2,0} = 0,\]
\[\gamma_0 c_{1,0} = -[(1-\lambda v_k)+\lambda v_k \cos(\xi h)]-\frac{1}{\tau}(\i\lambda v_k \sin(\xi h))^2,\]
\[ \gamma_0 d_{2,0}=\frac{1}{\tau}[(1-\lambda v_k)+\lambda v_k \cos(\xi h)]\i\lambda v_k\sin(\xi h)
    + \i\lambda v_k \sin(\xi h),\]
and hence
\[\tilde{\bb{L}}_0 =
    \begin{bmatrix}
    I_N \otimes I_{N_t}     &    O      \\
    O      &    I_N \otimes I_{N_t}    \\
    \end{bmatrix} +  \begin{bmatrix}
    \gamma_0 c_{1,0} W \otimes P    &    O      \\
    \tau \gamma_0 d_{2,0} W  \otimes P    &    O   \\
    \end{bmatrix}.\]

For $\gamma_0 c_{1,0}$, one easily gets
\[
\begin{aligned}
|\gamma_0 c_{1,0}|
&= \Big|-[(1-\lambda v_k)+\lambda v_k \cos(\xi h)]-\frac{1}{\tau}(\i\lambda v_k \sin(\xi h))^2 \Big|\\
&= \Big|1-\lambda v_k - \frac{\tau}{h^2}(v_k \sin(\xi h))^2 +  \lambda v_k \cos(\xi h)\Big|
=: |a + b|,
\end{aligned}
\]
where
\[a = 1-\lambda v_k - \frac{\tau}{h^2}(v_k \sin(\xi h))^2 , \qquad b = \lambda v_k \cos(\xi h).\]
Noting that
\[a = 1-\lambda v_k - \frac{\tau}{h^2}(v_k \sin(\xi h))^2 \ge 1-\lambda- \frac{\tau}{h^2},\]
one has $a\ge 0$ when
\begin{equation}\label{cflth}
\lambda + \frac{\tau}{h^2} \le 1  \quad \mbox{or} \quad  \frac{\tau}{h^2} \le \frac{1}{1+h}.
\end{equation}
Then,
\begin{align*}
|\gamma_0 c_{1,0}| = |a + b| \le |a|+|b|
= 1- \lambda v_k (1 -|\cos(\xi h)|)  - \frac{\tau}{h^2}(v_k \sin(\xi h))^2  = :b^{\prime},
\end{align*}
where the right-hand side satisfies $b^{\prime} \ge a \ge 0$ under the condition of \eqref{cflth}, which also implies $b^{\prime}\le 1$. Let $c =\lambda v_k ( 1 - \cos  (\xi h) )$. One has
\begin{align*}
|\tau \gamma_0 d_{2,0} |
& = | (1 + \tau) - \lambda v_k ( 1 - \cos  (\xi h) )| \cdot |\lambda v_k\sin(\xi h)| \\
& \le | (1 + \tau) - c | \le \max\{  |1 + \tau - c_{\min}|,   |1 + \tau - c_{\max}| \}\\
& = \max\{  | 1 + \tau + 0|,   |1 + \tau - 2| \} = 1 + \tau.
\end{align*}

3) We first consider the maximum singular value. Since $w_1 + \cdots + w_N = 1$, one can check that $WW^T = \|\bb{w}\|^2 \cdot  \bb{1}_N$, where $\|\bb{w}\|^2 = w_1^2 + \cdots + w_N^2$, and $\bb{1}_N$ is the $N$-th order matrix with all entries being 1. Then,
\[\|WW^T\| \le N \|\bb{w}\|^2 \le N (w_1 + \cdots + w_N)^2 = N, \]
which gives $\|W\| \le N^{1/2}$ and
\begin{align*}
\sigma_{\max}(\tilde{\bb{L}}_0) =  \|\tilde{\bb{L}}_0\|
 \le 1 + \max\{\gamma_0 c_{1,0}, \tau \gamma_0 d_{2,0} \} \|W\| \cdot \|P\|  \lesssim N^{1/2}.
\end{align*}

4) For the minimum singular value, since $\sigma_{\min }(\tilde{\bb{L}}_0) = 1/\|\tilde{\bb{L}}_0^{-1}\|$, we only need to provide a upper bound for $\|\tilde{\bb{L}}_0^{-1}\|$. By definition,
\[
\|\tilde{\bb{L}}_0^{-1}\| = \max_{\| \bb{b}\| \le 1}\|\tilde{\bb{L}}_0^{-1} \bb{b}\|, \qquad \bb{b} = [\bb{f}; \bb{g}],  \]
where $\tilde{\bb{L}}_0^{-1} \bb{b}$ clearly corresponds to the following linear system
\begin{equation*}
\begin{cases}
\tilde{r}_k^{n+1} + \gamma_0 c_{1,0}\sum_{k'=1}^{N}w_{k'}\tilde{r}_{k'}^n= \bb{f}_k^n, \\
\tilde{j}_k^{n+1} + \tau \gamma_0 c_{2,0}\sum_{k'=1}^{N}w_{k'}\tilde{r}_{k'}^n= \bb{g}_k^n,
\end{cases}
\end{equation*}
which can be written in matrix form as
\begin{equation*}
\begin{cases}
\tilde{\bb{r}}^{n+1} = A \tilde{\bb{r}}^n +  \bb{f}^n,  \qquad A = - \gamma_0 c_{1,0} W,\\
\tilde{\bb{j}}^{n+1} = B \tilde{\bb{r}}^n + \bb{g}^n, \qquad B = - \tau \gamma_0 c_{2,0}W.
\end{cases}
\end{equation*}
Assume the maximum value is attained  at $\bb{b}$.  Since $|\gamma_0 c_{1,0}|\le 1$, $\|W\|\le N^{1/2}$ and $W^2 = W$,
\[
\|\tilde{\bb{r}}^n\| \le \|A^n\| \|\bb{f}^0\| + \|A^{n-1}\| \|\bb{f}^1\| + \cdots + \|\bb{f}^{n-1}\|
\le N^{1/2} (\|\bb{f}^0\| + 1).
\]
Similarly, we obtain from $\tau \gamma_0 c_{2,0} \le 1+\tau$ that
\[\|\tilde{\bb{j}}^n\| \le (1+\tau)^n N^{1/2} (\|\bb{g}^0\| + 1) \lesssim N^{1/2} (\|\bb{g}^0\| + 1).\]
Thus,
\[\|\tilde{\bb{L}}_0^{-1}\| =  \|\tilde{\bb{L}}_0^{-1} \bb{b} \| \lesssim N^{1/2}N_t (\tilde{\bb{r}}^0 + 1),\]
which shows $\sigma_{\min }(\tilde{\bb{L}}_0) \gtrsim 1/(N^{1/2}N_t)$.

5) Now we calculate the $L^2$ norm $\|\tilde{\bb{L}}_{\epsilon}-\tilde{\bb{L}}_{0}\|_{2}$ for the perturbation term, where
\[\tilde{\bb{L}}_{\epsilon}-\tilde{\bb{L}}_{0}=
    \begin{bmatrix}
    I_N \otimes c_{1,\epsilon}P + (\gamma c_{1,\epsilon}-\gamma_0 c_{1,0}) W \otimes P    &    \tau^{-1} I_N \otimes (c_{2,\epsilon}P)      \\
    \tau ( I_N \otimes d_{2,\epsilon}P +  (\gamma d_{2,\epsilon}-\gamma_0 d_{2,0}) W  \otimes P)    &    I_N \otimes d_{1,\epsilon}P  \\
    \end{bmatrix}.\]
A straightforward calculation gives
\[
\begin{cases}
 |c_{1,\epsilon}| \lesssim \frac{\epsilon^2}{\tau+\epsilon^2}
+\frac{\epsilon^2(1-\epsilon^2)}{(\epsilon^2+\tau)^2}, \\
|\gamma c_{1,\epsilon}-\gamma_0 c_{1,0}|
\lesssim \frac{\epsilon^2}{\tau+\epsilon^2}+\frac{\epsilon^2(\epsilon^2+2\tau+\tau^2)}{\tau(\epsilon^2+\tau)^2},\\
|\tau^{-1} c_{2,\epsilon}|\lesssim
\frac{\epsilon^2}{\tau(\epsilon^2+\tau)},\\
|\tau d_{2,\epsilon}| \lesssim
\frac{\tau (1-\epsilon^2)\epsilon^2}{(\epsilon^2+\tau)^2}
    +\frac{\tau \epsilon^2}{\epsilon^2+\tau},\\
|\tau (\gamma d_{2,\epsilon}-\gamma_0 d_{2,0})| \lesssim
    \frac{\epsilon^2(\epsilon^2+2\tau+\tau^2)}{(\epsilon^2+\tau)^2}
    +\frac{\tau \epsilon^2}{\epsilon^2+\tau},\\
|d_{1,\epsilon}|\lesssim \frac{\epsilon^2}{\tau+\epsilon^2}.
\end{cases}
\]
Since $\|W\|\le N^{1/2}$ and $\|P\|\le 1$,  there holds
\begin{equation*}
    \begin{cases}
    \|I_N \otimes c_{1,\epsilon}P\| \le \|I_N\|\|c_{1,\epsilon}P\|
    \le |c_{1,\epsilon}|, \\
    \|(\gamma c_{1,\epsilon}-\gamma_0 c_{1,0}) W \otimes P\|\le
    (\gamma c_{1,\epsilon}-\gamma_0 c_{1,0})\|W\|\|P\|
    \le N^{1/2} |\gamma c_{1,\epsilon}-\gamma_0 c_{1,0}|, \\
    \|\tau^{-1} ( I_N \otimes c_{2,\epsilon}P)\|\le
    \tau^{-1} c_{2,\epsilon} \|I_N\|\|P\|
    \le |\tau^{-1} c_{2,\epsilon}|, \\
    \|\tau I_N \otimes d_{2,\epsilon}P\|\le
    \tau d_{2,\epsilon} \|I_N\|\|P\|
    \le |\tau d_{2,\epsilon}|\\
    \|\tau (\gamma d_{2,\epsilon}-\gamma_0 d_{2,0}) W  \otimes P)\|
    \le \tau (\gamma d_{2,\epsilon}-\gamma_0 d_{2,0})\|W\|\|P\|
    \le N^{1/2}|\tau (\gamma d_{2,\epsilon}-\gamma_0 d_{2,0})|, \\
    \|I_N \otimes d_{1,\epsilon}P\|\le
    d_{1,\epsilon}\|I_N \|\|P\|
    \le |d_{1,\epsilon}|,
     \end{cases}
\end{equation*}
and
\begin{equation*}
    \begin{cases}
    \|\tilde{\bb{L}}_{11,\epsilon}-\tilde{\bb{L}}_{11,0}\|
    \le |c_{1,\epsilon}|+N^{1/2}|\gamma c_{1,\epsilon}-\gamma_0 c_{1,0}|, \\
    \|\tilde{\bb{L}}_{12,\epsilon}-\tilde{\bb{L}}_{12,0}\|
    \le |\tau^{-1} c_{2,\epsilon}|, \\
    \|\tilde{\bb{L}}_{21,\epsilon}-\tilde{\bb{L}}_{21,0}\|
    \le |\tau d_{2,\epsilon}|+N^{1/2}|\tau (\gamma d_{2,\epsilon}-\gamma_0 d_{2,0})|, \\
    \|\tilde{\bb{L}}_{22,\epsilon}-\tilde{\bb{L}}_{22,0}\|
    \le |d_{1,\epsilon}|, \\
     \end{cases}
\end{equation*}
This means $\| \tilde{\bb{L}}_{\epsilon}-\tilde{\bb{L}}_{0}\| \le \alpha(\epsilon)$, with
\[
\alpha(\epsilon)=\frac{\epsilon^2}{\tau+\epsilon^2}(N^{1/2}\tau+N^{1/2}+\tau+\frac{1}{\tau})
+\frac{\epsilon^2(1-\epsilon^2)}{(\epsilon^2+\tau)^2}(1+\tau)
+\frac{\epsilon^2(\epsilon^2+2\tau+\tau^2)}{\tau(\epsilon^2+\tau)^2}N^{1/2}(1+\frac{1}{\tau}).
\]

6) Finally we analyze the time complexity.
The classical algorithm is to iteratively solve the following equations
\begin{equation*}
\begin{cases}
\bb{r}^{n+1}=\bb{B}_1\bb{r}^n - \bb{A}_1 \bb{j}^n + \widetilde{\bb{f}}^{n+1}\\
\bb{j}^{n+1}=\bb{A}_2 \bb{j}^n - \bb{B}_2 \bb{r}^n + \widetilde{\bb{g}}^{n+1}.
\end{cases}
\end{equation*}
The sparsity of $\bb{B}_i$ and $\bb{A}_i$ is $\mathcal{O}(N)$ and the matrix order is $\mathcal{O}(NN_x)$. Thus
the time complexity of each iteration step is $\mathcal{O}(N^2N_x)$, and the time complexity after $N_t$ iterations is
\[C = \mathcal{O}(N^2 N_t N_x).\]

For the quantum treatment, by the estimates of singular values, the condition number $\kappa=\mathcal{O}(N N_t)$, and the sparsity $s=\mathcal{O}(N)$. Under the given conditions, the error bound $\delta = \mathcal{O}(h)$. Plug these quantities in \eqref{cpCAS} to get
\[ Q=\mathcal{O}(s\kappa \log(1/\delta)) = \mathcal{O}( N^2 N_t \log N_x). \]
This completes the proof.
\end{proof}

\section{Time complexity of the explicit scheme}

As a comparison, in this section we discuss the time complexity of the explicit scheme for both the classical and quantum treatments.
We use the upwind finite difference to discretize \eqref{equ:1}. The upwind scheme is
\begin{align}
 \frac{f_{k,m}^{n+1}-f_{k,m}^n}{\tau}
+\frac{1}{\epsilon}v_k^+ \frac{f_{k,m}^n-f_{k,m-1}^n}{h}+\frac{1}{\epsilon}v_k^-\frac{f_{k,m+1}^n-f_{k,m}^n}{h}
 =\frac{1}{\epsilon^2}\Big(\frac{1}{2}\sum_{k'=-N}^{N}w_{k'}f_{k',m}^n-f_{k,m}^n\Big), \label{equ:4_1}
\end{align}
or
\[f_{k,m}^{n+1}- \Big[ 1 - \frac{\lambda}{\epsilon} ( v_k^+ -v_k^- ) - \frac{\tau}{\epsilon^2} \Big]  f_{k,m}^n
-\frac{\lambda}{\epsilon}v_k^+ f_{k,m-1}^n + \frac{\lambda}{\epsilon} v_k^- f_{k,m+1}^n - \frac{\tau}{2\epsilon^2} \sum_{k'}w_{k'}f_{k',m}^n = 0. \]
Let~$\bb{f}_m=[f_{-N,m},f_{-N+1,m},f_{-N+2,m},\cdots,f_{N-2,m},f_{N-1,m},f_{N,m}]^T$ and define
\[
\lambda = \frac{\tau}{\epsilon},\qquad  c_k = 1 - \frac{\lambda}{\epsilon} ( v_k^+ -v_k^- ) - \frac{\tau}{\epsilon^2}.
\]
Then \eqref{equ:4_1} can be written as
\begin{equation}
\label{equ:4_2}
\bb{f}_m^{n+1}-\bb{C}\bb{f}_m^n
-\frac{\lambda}{\epsilon}\bb{V}^+\bb{f}_{m-1}^n
+\frac{\lambda}{\epsilon}\bb{V}^-\bb{f}_{m+1}^n
-\frac{\tau}{2\epsilon^2}\bb{W}\bb{f}_m^n
= \bb{0},
\end{equation}
where,
\[\bb{C} = \text{diag}(c_{-N},\cdots, c_{-1}, c_1, \cdots, c_N), \qquad
\bb{V}^\pm = \text{diag}(v_{-N}^\pm,\cdots, v_{-1}^\pm, v_1^\pm, \cdots, v_N^\pm),\]
\[\bb{W} = \begin{bmatrix}
w_{-N}      &   \cdots  & w_{-1}    &  w_1  &  \cdots  &  w_N   \\
\vdots      &    \vdots      & \vdots  &\vdots   &  \vdots & \vdots\\
w_{-N}      &   \cdots  & w_{-1}    &  w_1  &  \cdots  &  w_N   \\
\end{bmatrix}_{2N \times 2N}.
\]
Let $\bb{f}^n=[\bb{f}_1^n; \bb{f}_2^n; \cdots;\bb{f}_{N_x}^n]$. One has
\begin{equation}
    \bb{f}^{n+1}-\bb{B}\bb{f}^n = \bb{b}^n,
\end{equation}
where $\bb{b}^n = \frac{\lambda}{\epsilon} [\bb{V}^+\bb{f}_0^n; \bb{0}; \cdots; \bb{0}; -\bb{V}^-\bb{f}_{N_x+1}^n]$ and
\[
\bb{B}=
\begin{bmatrix}
\bb{C}+\frac{\tau}{2\epsilon^2} \bb{W}&-\frac{\lambda}{\epsilon} \bb{V}^-&  &&\\
\frac{\lambda}{\epsilon}\bb{V}^+&\bb{C}+\frac{\tau}{2\epsilon^2}\bb{W}&-\frac{\lambda}{\epsilon}\bb{V}^-& &  \\
&\ddots& \ddots  &\ddots     &\\
&&\frac{\lambda}{\epsilon}\bb{V}^+&\bb{C}+\frac{\tau}{2\epsilon^2}\bb{W}&-\frac{\lambda}{\epsilon}\bb{V}^-\\
&&&\frac{\lambda}{\epsilon}\bb{V}^+&\bb{C}+\frac{\tau}{2\epsilon^2} \bb{W}\\
\end{bmatrix}.\]
Let $\bb{U}=[\bb{f}^1; \bb{f}^2; \cdots; \bb{f}^{N_t}]$. The linear system for the quantum difference approach can be written as
\begin{equation}
\label{equ:4-4}
    \bb{L}\bb{U}=\bb{F},
\end{equation}
where,
\[
\bb{L}=
    \begin{bmatrix}
    \bb{I}    &           &                &          \\
    -\bb{B}   &  \bb{I}   &                &          \\
              & \ddots    &   \ddots       &          \\
              &           &     -\bb{B}    & \bb{I}   \\
    \end{bmatrix}, \qquad
\bb{F}=
    \begin{bmatrix}
    \bb{b}^0+\bb{B}\bb{f}^0\\
    \bb{b}^1\\
    \vdots\\
    \bb{b}^{N_t-1},
    \end{bmatrix}.
\]

\begin{theorem}\label{the:4.1}
Let $\delta$ be the error bound and let $h = \mathcal{O}(\epsilon \delta)$ be the spatial step. Suppose the temporal step satisfies $\tau \le \frac{h}{\epsilon+h} \epsilon^2$. Then one has
\begin{enumerate}[(1)]
  \item The singular values of the coefficient matrix in \eqref{equ:4-4} satisfies
\[\sigma_{\min} \gtrsim 1/(N^{1/2} N_t) , \qquad \sigma_{\max} \lesssim N^{1/2}.\]
  \item The condition number $\kappa = \mathcal{O}(NN_t)$ and the sparsity $s = \mathcal{O}(N)$.
  \item  The time complexities of the classical treatment and the quantum treatment for solving~\eqref{equ:4-4}~are
\[C = \mathcal{O}(N^2 N_t N_x) = \mathcal{O}(N^2 \epsilon^{-3} \delta^{-1}), \]
\[Q = \mathcal{O}(N^2 N_t \log N_x) = \mathcal{O}(N^2 \epsilon^{-2} \log((\epsilon \delta)^{-1})). \]
\end{enumerate}
\end{theorem}
\begin{proof}
1) The truncation error is $\mathcal{O}(\tau + h/\epsilon + ((2N)!)^{-1}/\epsilon^2)$.
Let the error bound be $\delta$. Then we can choose $h = \mathcal{O}(\epsilon \delta)$.

2) We first consider the minimum singular value. By definition, one has $\sigma_{\min}(\bb{L}) = 1/\|\bb{L}^{-1}\|$.
It suffices to give an upper bound of $\|\bb{L}^{-1}\|$. By a direct calculation, one obtains
\begin{equation*}
\bb{L}^{-1}=
\begin{bmatrix}
        \bb{I} & & & & \\
        \bb{B} & \bb{I} & & & \\
        \bb{B}^{2} & \ddots & \ddots & & \\
        \vdots & \ddots & \ddots & \ddots & \\
        \bb{B}^{N_{t}-1} & \cdots & \bb{B}^{2} & \bb{B} & \bb{I}
\end{bmatrix}
=
\begin{bmatrix}
        \bb{I} & & & & \\
          & \bb{I} & & & \\
        &   & \ddots & & \\
        & &   & \ddots & \\
        & & &   & \bb{I}
\end{bmatrix}
+
\begin{bmatrix}
        \bb{O} & & & & \\
        \bb{B} &\bb{O} & & & \\
        & \ddots &\ddots   & & \\
        & & \ddots & \ddots  & \\
        & & & \bb{B} &\bb{O}
\end{bmatrix}
+
\cdots,
\end{equation*}
hence,
\[\|\bb{L}^{-1}\| \le \|\bb{I}\|+\|\bb{B}\|+\cdots+ \|\bb{B}^{N_t-1} \|.\]

Now we estimate~$\|\bb{B}^n\|$.
The matrix $\bb{B}$ can be split as $\bb{B} = \bb{B}_1 + \alpha \bb{B}_2$, where $\alpha = \tau/\epsilon^2$,
\[
\bb{B}_1=
\begin{bmatrix}
\bb{C}&-\frac{\lambda}{\epsilon} \bb{V}^-&  &&\\
\frac{\lambda}{\epsilon}\bb{V}^+&\bb{C}&-\frac{\lambda}{\epsilon}\bb{V}^-& &  \\
&\ddots& \ddots  &\ddots     &\\
&&\frac{\lambda}{\epsilon}\bb{V}^+&\bb{C}&-\frac{\lambda}{\epsilon}\bb{V}^-\\
&&&\frac{\lambda}{\epsilon}\bb{V}^+&\bb{C}\\
\end{bmatrix}, \quad
\bb{B}_2=
\frac{1}{2}\begin{bmatrix}
 \bb{W}& &  &&\\
 & \bb{W}& & &  \\
&& \ddots  &     &\\
&& & \bb{W}&\\
&&& & \bb{W}\\
\end{bmatrix}.\]
When $\tau \le \frac{h}{\epsilon+h} \epsilon^2$~or~$\lambda \le \epsilon^2/(\epsilon+h)$, one easily finds that
\[ c_k = 1 - \frac{\lambda}{\epsilon} ( v_k^+ -v_k^- ) - \frac{\tau}{\epsilon^2} \ge 0. \]
Since $v_k^+\ge 0$ and $v_k^- \le  0$, using the Gershgorin-type lemma for singular values, one gets
\[\|\bb{B}_1\| \le  \Big( 1 - \frac{\lambda}{\epsilon} ( v_k^+ -v_k^- ) - \frac{\tau}{\epsilon^2}\Big)
+ \frac{\lambda}{\epsilon}v_k^+ - \frac{\lambda}{\epsilon} v_k^- = 1 - \frac{\tau}{\epsilon^2}, \]
and $\|\bb{B}_1\|+\alpha\le 1$.  It is easy to check that $\bb{B}_2^2 = \bb{B}_2$, yielding
\[\bb{B}^n
 = (\bb{B}_1 + \alpha \bb{B}_2)^n = \sum\limits_{k = 0}^n   C_n^k \bb{B}_1^{n-k} (\alpha \bb{B}_2)^k
= \bb{B}_2 \sum\limits_{k = 0}^n   C_n^k \bb{B}_1^{n-k} \alpha^k,\]
and hence,
\[\|\bb{B}^n\|
 \le \|\bb{B}_2\| \sum\limits_{k = 0}^n   C_n^k \|\bb{B}_1\|^{n-k} \alpha^k
 =\frac{1}{2}\|\bb{W}\|(\|\bb{B}_1\|+\alpha)^n \le \frac12 \|\bb{W}\|
 \lesssim N^{1/2}.
\]
This shows
 \[\|\bb{L}^{-1}\| \lesssim N^{1/2} N_t \qquad \mbox{or} \qquad  \sigma_{\min}(\bb{L}) \gtrsim 1/(N^{1/2} N_t).\]

 3) For the maximum singular value, one obtains from the Gershgorin-type lemma that
\[
\|\bb{L}\|\le 1+\|\bb{B}\| \le 1 + \|\bb{B}_1\| + \alpha \|\bb{B}_2\| \le 2 + \frac12 \|\bb{W}\| \lesssim N^{1/2}.
\]

 4) It is obvious that the sparsity $s = \mathcal{O}(N)$. From the above estimates,  we know that the condition number  $\kappa = \mathcal{O}(NN_t)$.

 5) The analysis of the time complexity can be carried out as that of the diffusive relaxation scheme.
\end{proof}

\section{Conclusions}

We studied the time complexities of finite difference methods for solving the multiscale transport equation in the setting of quantum computing.
Our results show that the quantum implementation of the classical Asymptotic-Preserving schemes, a popular multiscale framework for multiscale problems \cite{Jin2022Acta}~--~ is equally important in quantum computing since they allow the computational costs for quantum algorithms to be {\it independent} of the small physical scaling parameters. This study also suggests that one should take full advantage of state-of-the-art multiscale classical algorithms when designing quantum algorithms for multiscale PDEs.

\subsection*{Declaration of competing interest}

The authors declare that they have no known competing financial interests or personal relationships that could have
appeared to influence the work reported in this paper.

\section*{Acknowledgement}
SJ was partially supported by the NSFC grant No.~12031013, the Strategic Priority Research Program of Chinese Academy of Sciences
XDA25010404, the Shanghai Municipal Science and Technology Major Project (2021SHZDZX0102), and  the Innovation Program of Shanghai Municipal Education Commission (No. 2021-01-07-00-02-E00087).  YY was partially supported by China Postdoctoral Science Foundation (no. 2022M712080).

\bibliographystyle{plain} 
\bibliography{Refs}

\begin{thebibliography}{10}

\bibitem{Berry-2014}
D.~W. Berry.
\newblock High-order quantum algorithm for solving linear differential
  equations.
\newblock {\em J. Phys. A: Math. Theor.}, 47(10):105301, 17 pp., 2014.

\bibitem{BerryChilds2017ODE}
D.~W. Berry, A.~M. Childs, A.~Ostrander, and G.~Wang.
\newblock Quantum algorithm for linear differential equations with
  exponentially improved dependence on precision.
\newblock {\em Comm. Math. Phys.}, 356(3):1057--1081, 2017.

\bibitem{CaseZweifel}
K.~W. Case and P.~F. Zweifel.
\newblock {\em Linear Transport Theory}.
\newblock Addison-Wesley, Reading, MA, 1967.

\bibitem{Chander}
S.~Chandrasekhar.
\newblock {\em Radiative Transfer}.
\newblock Dover, New York, 1960.

\bibitem{Childs-2017}
A.~M. Childs, R.~Kothari, and R.~D. Somma.
\newblock Quantum algorithm for systems of linear equations with exponentially
  improved dependence on precision.
\newblock {\em SIAM J. Comput.}, 46(6):1920--1950, 2017.

\bibitem{Childs-2021}
A.~M. Childs, J.~P. Liu, and A.~Ostrander.
\newblock High-precision quantum algorithms for partial differential equations.
\newblock {\em Quantum}, 5:574, 2021.

\bibitem{Costa2021QLSA}
P.~C.~S. Costa, D.~An, Y.~A. Sanders, Y.~Su, R.~Babbush, and D.~W. Berry.
\newblock Optimal scaling quantum linear systems solver via discrete adiabatic
  theorem.
\newblock {\em arXiv:2111.08152}, 2021.

\bibitem{Dobin2021plasma}
I.~Y. Dobin and E.~A. Startsev.
\newblock On applications of quantum computing to plasma simulations.
\newblock {\em Phys. Plasmas}, 28:092101, 2021.

\bibitem{Gilyen2019QSVD}
A.~Gily\'{e}n, Y.~Su, G.~Low, and N.~Wiebe.
\newblock Quantum singular value transformation and beyond: exponential
  improvements for quantum matrix arithmetics.
\newblock In {\em Proceedings of the 51st Annual ACM SIGACT Symposium on Theory
  of Computing}, pages 193--204. STOC, 2019.

\bibitem{gleinig2021efficient}
N.~Gleinig and T.~Hoefler.
\newblock An efficient algorithm for sparse quantum state preparation.
\newblock {\em 2021 58th ACM/IEEE Design Automation Conference (DAC)}, pages
  433--438, 2021.

\bibitem{HHL-2009}
A.~W. Harrow, A.~Hassidim, and S.~Lloyd.
\newblock Quantum algorithm for linear systems of equations.
\newblock {\em Phys. Rev. Lett.}, 103(15):150502, 4 pp., 2009.

\bibitem{Jin2022Acta}
S.~Jin.
\newblock Asymptotic-preserving schemes for multiscale physical problems.
\newblock {\em Acta Numer.}, 31:415--489, 2022.

\bibitem{JinLiu2022nonlinear}
S.~Jin and N.~Liu.
\newblock Quantum algorithms for computing observables of nonlinear partial
  differential equations.
\newblock {\em arXiv:2202.07834}, 2022.

\bibitem{JinLiuYu-2}
S.~Jin, N.~Liu, and Y.~Yu.
\newblock Time complexity analysis of quantum algorithms via linear
  representations for nonlinear ordinary and partial differential equations.
\newblock {\em arXiv:2209.08478}, 2022.

\bibitem{JLY2022multiscale}
S.~Jin, N.~Liu, and Y.~Yu.
\newblock Time complexity analysis of quantum difference methods for linear
  high dimensional and multiscale partial differential equations.
\newblock {\em J. Comput. Phys.}, 471:111641, 2022.

\bibitem{JPT1998}
S.~Jin, L.~Pareschi, and G.~Toscani.
\newblock Diffusive relaxation schemes for multiscale discrete-velocity kinetic
  equations.
\newblock {\em SIAM J. Numer. Anal.}, 35(6):2405--2439, 1998.

\bibitem{Jin-Pareschi-Toscani-2000}
Shi Jin, Lorenzo Pareschi, and Giuseppe Toscani.
\newblock Uniformly accurate diffusive relaxation schemes for multiscale
  transport equations.
\newblock {\em SIAM J. Numer. Anal.}, 38:913--936, 2000.

\bibitem{Joseph2020KvN}
I.~Joseph.
\newblock Koopman--von {N}eumann approach to quantum simulation of nonlinear
  classical dynamics.
\newblock {\em Phys. Rev. Research}, 2(4):043102, 2020.

\bibitem{Lin-Montanaro-Shao-2020}
N.~Linden, A.~Montanaro, and C.~Shao.
\newblock Quantum vs. classical algorithms for solving the heat equation.
\newblock {\em arXiv:2004.06516}, 2020.

\bibitem{Liu2021nonlinear}
J.~Liu, H.~{\O}. Kolden, H.~K. Krovi, N.~F. Loureiro, K.~Trivisa, and A.~M.
  Childs.
\newblock Efficient quantum algorithm for dissipative nonlinear differential
  equations.
\newblock {\em Proc. Natl. Acad. Sci. U. S. A.}, 118(35), 2021.

\bibitem{Lloyd2020nonlinear}
S.~Lloyd, G.~D. Palma, C.~Gokler, B.~Kiani, Z.~Liu, M.~Marvian, F.~Tennie, and
  T.~Palmer.
\newblock Quantum algorithm for nonlinear differential equations.
\newblock {\em arXiv:2011.06571}, 2020.

\bibitem{Low-2019}
G.~H. Low.
\newblock Hamiltonian simulation with nearly optimal dependence on spectral
  norm.
\newblock In {\em Proceedings of the 51st Annual ACM SIGACT Symposium on Theory
  of Computing}, pages 491--502. STOC, 2019.

\bibitem{MRS}
P.~A. Markowich, C.~A. Ringhofer, and C.~Schmeiser.
\newblock {\em Semiconductor Equations}.
\newblock Springer, Vienna, 1990.

\bibitem{Ryzhik}
L.~Ryzhik, G.~Papanicolaou, and J.~B. Keller.
\newblock Transport equations for elastic and other waves in random media.
\newblock {\em Wave Motion}, 24(4):327--370, 1996.

\bibitem{Stewart1990PerturbationTF}
G.~W. Stewart.
\newblock Perturbation theory for the singular value decomposition.
\newblock In {\em SVD and Signal Processing, II. Algorithms, Analysis and
  Applications}, pages 99--109. Elsevier, 1990.

\bibitem{zhang2022quantum}
X.~Zhang, T.~Li, and X.~Yuan.
\newblock Quantum state preparation with optimal circuit depth: Implementations
  and applications.
\newblock {\em arXiv:2201.11495}, 2022.

\end{thebibliography}

\end{CJK*}

\end{document}